\documentclass{article}

\input{tcilatex}
\begin{document}

\begin{center}
\textbf{Inertia in the Structure of Four-dimensional Space}

\textsl{A.M. Gevorkian\footnote{%
e-mail: dealing51@hotmail.com} , R. A. Gevorkian* }

\textsl{Institute of Physical Research, 378410, Ashtarak-2, Republic of
Armenia}

\textsl{*Institute of Radiophisics and Electronics, 378410, Ashtarak-2,
Republic of Armenia}

\smallskip

\textbf{Abstract}
\end{center}

\textit{1. Following Rimman, Minkowski and Einstein, for the first time
equations of the inert filed in the covariant form are found geometrically.}

\textit{2. In the approximation of a weak field for the first time the Law
of Inertia in a material space (as opposed to the absolute space) is
received. A consequence is the formulation of Mach's principle.}

\textit{3. Analogous to Einstein's expression for the gravitational field }$%
c=c_0\left( 1-\dfrac{2GM}r\right) ,$\textit{\ Compton's formula is received
for the inert field }$r=\dfrac h{mc}\left( 1-l_0\cos \theta \right) $\textit{%
.}

\textit{4. For the first time transcendental equations are received, one of
the solutions of which corresponds to the value of the magnetic charge by
Dirack, or to the constant of fine structure.}

\smallskip

Minkowski's work [1] for the first time led the solution of physical
problems to geometrical problems.

1. Such physical theorem as kinematics and theory of inertial systems is
obtained on the basis of flat space through equation

\begin{equation}
R_{iklm}=0  \label{eq1}
\end{equation}

for three-dimensional Euclidean as well as for four-dimensional
pseudo-Euclidean spaces.

2.

\begin{equation}
G_{ik}=R_{ik}-\dfrac 12Rg_{ik}  \label{eq2}
\end{equation}

is presented as an object of four-dimensional pseudo-Rimman geometry
(Einstein) and is determined by formula

\begin{equation}
K=\dfrac{G_{ik}\xi ^i\xi ^k}{\xi ^i\xi _i}  \label{eq3}
\end{equation}

where $K$ is the mean scalar crookedness of three-dimensional subspace,
which is orthogonal to an arbitrarily determined vector $\xi ^i$. Tensor $%
G_{ik}$ is called the conservative tensor of Gilbert-Einstein and plays a
fundamental role in the structure of the field equation of the General
Theory of Relativity (GTR).

Let us discuss formula

\begin{equation}
G_{ik}=0  \label{eq4}
\end{equation}

Einstein [2] predicted that this equation might be a basis for a more
general physical theory in case it described electromagnetic fields along
with gravitational fields. It is true that this equation is insufficient for
describing all physical fields. It describes only gravitational fields
outside of the source. From the point of view of geometry this equation is a
necessary but not sufficient condition for flat space.

For a precise description of the flat space along with equation (4) we will
try to introduce one more geometric equation. The starting point of the
unified field theory\footnote{%
Inert field is included with the fieled of gravitation} is the fact that
equation (4) alone is insufficient for describing all phenomena. Therefore
the problem of finding a mathematical object occurs which, after being added
to $G_{ik}$ , will allow to obtain the structure of space describing the
physical situation completely.

Gravitation can be described either by introducing the gravitational field
into the flat space, or by discussing a geometrical object in crooked
four-dimensional space without the gravitational field. The last simulation
is the geometric interpretation of gravitation. Hence if the deformation,
which creates crookedness is excluded, we will obtain the flat space.

\smallskip

\emph{Mathematical Part}

Let us introduce tensor $\overline{G}_{ik}$, $P_{ik}$ in the following way:

\[
G_{ik}+\dfrac{\left( \eta _{ik}-g_{ik}\right) }{\xi ^i\xi _i}=P_{ik} 
\]

\begin{equation}
\overline{G}_{ik}-\dfrac{\left( \eta _{ik}-g_{ik}\right) }{\xi ^i\xi _i}%
=P_{ik}  \label{eq5}
\end{equation}

$\overline{G}_{ik}$ is the counter-tensor of $G_{ik}$ , which describes the
counter-space in relation to space determined by $G_{ik}$. $\eta _{ik}$, $%
g_{ik}$ are metric tensors of flat and crooked spaces respectively, and $\xi
^i$ is the vector determined initially.

Geometrically (5) means that two spaces of Einstein $\overline{G}$ and $G$
are called mutually counter if the following is realized: In each point of
space the ratio of mean scalar crookednesses of the three-dimensional
subspace orthogonal to initially determined vector $\xi ^i$ depends on the
module of that vector only. Both spaces have a common coordination.

\begin{equation}
\dfrac K{\overline{K}}=\dfrac{e_ie^i}{\xi ^i\xi _i}  \label{eq6}
\end{equation}

where $e_i$ is the unit vector of vector $\xi _{i\text{.}}$

The mean scalar crookedness of the three-dimensional subspace is determined
by formula (3) and therefore (6) is written as

\begin{equation}
\xi ^i\xi _iG_{ik}\xi ^i\xi ^k=e_ie^i\overline{G}_{ik}\xi ^i\xi ^k
\label{eq7}
\end{equation}

Hence we can determine

\[
G_{ik}-2\dfrac{g_{ik}}{\xi ^i\xi _i}=\overline{G}_{ik}-2\dfrac{\eta _{ik}}{%
\xi ^i\xi _i} 
\]

\begin{equation}
G_{ik}+\dfrac{\left( \eta _{ik}-g_{ik}\right) }{\xi ^i\xi _i}=\overline{G}%
_{ik}-\dfrac{\left( \eta _{ik}-g_{ik}\right) }{\xi ^i\xi _i}  \label{eq8}
\end{equation}

Let us introduce operator $\left( -\right) ^n$ which transforms space into
counter-space, tensor $G_{ik}$ into tensor $\overline{G}_{ik}$, and the sign
is changed into the opposite one. Even times of counter-operating transforms
the space into itself, and odd times of counter-operating transforms the
space into the counter-space. If the left side is noted as tensor $P_{ik}$
and the right side is noted as counter-tensor $\left( -1\right) P_{ik}$,
then equation (8) can be written as

\begin{equation}
P_{ik}=\left( -1\right) P_{ik}  \label{eq9}
\end{equation}

\smallskip \textit{Theorem 1}

For the space to be flat it is necessary and sufficient that tensor $G_{ik}$
and counter-tensor $\overline{G}_{ik}$ be equal.

\textit{Proof} - \textit{necessity}: From equation system (5) we receive

\begin{equation}
G_{ik}-\overline{G}_{ik}-2\dfrac{\left( \eta _{ik}-g_{ik}\right) }{\xi ^i\xi
_i}=0  \label{eq10}
\end{equation}

\textit{Sufficiency}: If $G_{ik}-\overline{G}_{ik}=0$ is a flat space and
vice versa $\eta _{ik}-g_{ik}=0$, then $G_{ik}=\overline{G}_{ik}$. Theorem
proved.

From system (5) follows equation

\begin{equation}
G_{ik}+\overline{G}_{ik}=2P_{ik}  \label{eq11}
\end{equation}

If $\overline{G}_{ik}=0$, then we receive the already familiar structure of
Einstein equation [3], only if we substitute $2P_{ik}$ by tensor of energy
of impulse and tension.

\smallskip \textit{Theorem 2}

For the space to be flat, it is necessary and sufficient that $P_{ik}=0$.

Equation (5) is written as

\begin{equation}
G_{ik}+\dfrac{\left( \eta _{ik}-g_{ik}\right) }{\xi ^i\xi _i}=0  \label{eq12}
\end{equation}

Based on the example of spherical symmetry, equation (12) has a flat
solution (metrics of Minkowski).

\textit{Consequence}: For flat space we have

\[
G_{ik}-\overline{G}_{ik}=0 
\]

\[
G_{ik}+\overline{G}_{ik}=0 
\]

Hence, for the space to be flat, it is necessary and sufficient that

\begin{equation}
G_{ik}=0,\overline{G}_{ik}=0  \label{eq13}
\end{equation}

or that space and counter-space coincide and be equal to zero.

Thus the condition of sufficiency of $\overline{G}_{ik}=0$ is added to the
condition of necessity of flat space $G_{ik}=0$.

If we look at equation (11) from the mathematical point of view, then tensor 
$2P_{ik}$ is a source of crookedness or a source of deformation.

\smallskip

However!.. Expression (10) can be decomposed by two menas into two
equivalent equations:

\[
G_{ik}=2\dfrac{\left( \eta _{ik}-g_{ik}\right) }{\xi ^i\xi _i}\text{\qquad }%
\overline{G}_{ik}=0\text{ \qquad (a) \qquad (first way)} 
\]

\begin{equation}
\overline{G}_{ik}=-2\dfrac{\left( \eta _{ik}-g_{ik}\right) }{\xi ^i\xi _i}%
\text{ \qquad }G_{ik}=0\text{ \qquad (b) \qquad (second way)}  \label{eq14}
\end{equation}

\emph{Physical Part}

The first pair of equations (14,a) describes the internal problem of
gravitation, or the external problem of inertia, and the second pair of
equations (14,b) - vice versa. Based on the above, the full problem of
gravitation can be formulated:

\[
G_{ik}=0\text{ } 
\]

\begin{equation}
G_{ik}=2\dfrac{\left( \eta _{ik}-g_{ik}\right) }{\xi ^i\xi _i}  \label{eq15}
\end{equation}

where the first equation is the external problem, and the second equation is
the internal problem. In the right side of the second equation of system
(15), the tensor of energy, impulse and tension is substituted by $T_{ik}=2%
\dfrac{\left( \eta _{ik}-g_{ik}\right) }{\xi ^i\xi _i}$ - a tensor of
geometric origin (magnitude of deformation of space). The solution of system
(15) will be searched for spherical symmetry or for a centrally symmetric
field.

The solution of the first equation (15) is the famous solution of
Schwarzschild [4] and is expressed by metrics:

\begin{equation}
ds^2=\left( 1-\dfrac{r_{gr}}r\right) c^2dt^2-r^2\left( \sin ^2\theta
d\varphi ^2+d\theta ^2\right) -\dfrac{dr^2}{1-r_{gr}/r}  \label{eq16}
\end{equation}

and the second equation of system (15), which describes gravitational filed
inside the source, is expressed by the system of equations (is calculated by
components of the tensor):

\[
\exp \left( -\lambda \right) \left( \dfrac{\partial \lambda }{\partial r}%
\dfrac 1r-\dfrac 1{r^2}\right) -\dfrac 1{r^2}=0 
\]

\begin{equation}
\dfrac{\partial \lambda }{\partial t}=0  \label{eq17}
\end{equation}

where

\[
g_{00}=\exp \left( -\lambda \right) 
\]

\[
g_{33}=\exp \left( \lambda \right) 
\]

the solution of which is presented by metrics

\begin{equation}
ds^2=\left( 1-\dfrac r{r_k}\right) c^2dt^2-r^2\left( \sin ^2\theta d\varphi
^2+d\theta ^2\right) -\dfrac{dr^2}{1-r/r_k}  \label{eq18}
\end{equation}

For small distances it becomes the metrics of usual pseudo-Euclidean
geometry. Another constant is determined from the condition of coincidence
on the border of the internal and the external solutions. If in the solution
of Schwarzschild that constant is expressed as gravitational radius and is a
geometrical measure of the active gravitational mass, then for (18) that
constant represents the inert radius and expresses the geometrical measure
of the active inert mass. From the condition of coincidence on the border of
the internal and the external solutions, we have

\[
1-\dfrac{r_{gr}}r=1-\dfrac r{r_k},\text{ or \quad }r_{gr}r_k=r^2=r_0^2,\text{
\quad }\dfrac{Gm_{gr}}cr_k=r_0^2 
\]

If constant $r_0$ is the Plank distance (expressed through natural
constants), then for $r_k$ we receive expression

\[
r_k=\dfrac{
\rlap{\protect\rule[1.1ex]{.325em}{.1ex}}h%
}{mc} 
\]

Mass is introduced into physics with different notions: First as inert
resistance, inert passive mass $m_{PI}$, and secondly, as a constant of
association which shows how strongly the gravitational field $\varphi $
affects the body. This is a constant of association, which is the passive
gravitational mass $m_{PG}$ (as the mass of a test body). And finally, the
third notion is the active gravitational mass $m_{AG}$, as gravitational
charge or as intensity of the source of gravitational field. The first two
notions are included in the equation of motion:

\[
m_{PI}\dfrac{d^2x^i}{dt^2}=m_{PG}\dfrac{d\varphi }{dx^i} 
\]

The third notion is included in the gravitational potential. The principle
of equivalence assumes the universal law

\[
m_{in}=m_{gr,}\qquad m_{PI}=m_{PG}\text{ \qquad \qquad (c)} 
\]

Note that in formulas where metric coefficients represent potentials of
specific force fields, gravitational radius is included, which is determined
by the active gravitational mass ($m_{AG}$). In the solution of
Schwarzschild these are potentials of gravitational filed. Metrical
coefficients of the counter-space are interpreted as potentials of the inert
field.

Mass $m_{AI}$, included in the Compton radius, is the fourth notion of mass.
It is the center of dispersion of light and is included in the structure of
metrical tensor of the counter-space $\overline{g}_{ik}$. The fourth notion
of mass is presented as the active inert mass. This was noted by Siama. [5].

The equality between the active gravitational and the active inert masses,
which are present in formulas (16) and (18) respectively, is formulated as
the active principle of equivalence.

\[
m_{gr}=m_{in} 
\]

and as a consequence $r_0=\dfrac{\left( Gh\right) ^{1/2}}{c^3}$ is the
formula expressing the fundamental meaning of length, only if $%
r_{gr}=r_{in}=r_0$.

From the condition of coincidence on the border of the internal and the
external solutions, (16) and (18) coincide in point $r_0$ and are expressed
as

\begin{equation}
ds^2=\left( 1-\dfrac m{m_0}\right) c^2dt^2-r^2\left( \sin ^2\theta d\varphi
^2+d\theta ^2\right) -\dfrac{dr^2}{1-m/m_0}  \label{eq19}
\end{equation}

where $m_0=\left( \dfrac{
\rlap{\protect\rule[1.1ex]{.325em}{.1ex}}h%
c}G\right) ^{1/2}$ is the fundamental meaning of mass.

Thus if the metrics of Schwarzschild describes the external gravitational
filed with source $r_{gr}$, then metrics (18) describes the external inert
filed with source $r_k$.

Gravitational and inert fields are linked through the active principle of
equivalence in the counter-projection. Let us consider particular cases.

1. Approximation of a weak field

The expression of component $g_{00}$ of the metric tensor

\[
g_{00}=1-\dfrac r{r_k} 
\]

describes the potential of the filed, and the force is determined by formula

\begin{equation}
F=mc^2\dfrac{dg_{00}}{dr},\text{ \qquad }F=\dfrac{mc^2}{%
\rlap{\protect\rule[1.1ex]{.325em}{.1ex}}h%
/Mc}=\dfrac{Mmc^3}{
\rlap{\protect\rule[1.1ex]{.325em}{.1ex}}h%
}\dfrac{\vec{r}_0}{\left| r_0\right| }  \label{eq20}
\end{equation}

where $M$ is the active inert mass, $m$ is the mass of the test body, and $%
\vec{r}_0$ is the unit vector. In formula (20) force $F$ does not depend on
the distance, which means that in the formation of that force all components
and fragments of the Universe participate equally. And if the Universe is
homogeneous and isotropic then due to $M=0$ (effective mass), this force $%
F=0 $. Newton's first law is obtained - the theory of inertial systems. By
the way, in this approximation from Schwarzschild's solution the Law of
Terrestrial Gravitation is obtained.

The forces of inertia are so usual that it is very uncommon to think about
such a problem. Despite this, a lot has been written and spoken about this.
Here we would like to discuss sources of inertial forces. According to
Newton [6], the source of inertial forces is the absolute space. This means
that the absolute space can influence matter, and the opposite influence
form the matter is excluded. Thus, the interaction between the absolute
space and material bodies does not exist. This rather abstract point of view
corresponds to the interpretation of physical interactions on the basis of
the field theory. Therefore, a result of the proposed theory is that forces
of inertia impacting on the body depend on physical qualities of space and
all members filling the space.

Basically, Mach's principle is formulated: Inert qualities of an object are
determined by distribution of mass-energy in the whole space. On the basis
of Mach's principle and the active principle of equivalence, as well as on
the basis of Rimman's idea on geometry of space corresponding to physics and
playing an essential role in it, we were able to build this theory.

Rotation of a body relative to the system of static stars is equivalent to
the rotation of stars around the body. In both cases the relative motion is
the same. This is relativity according to Mach and Berkley.

Let us imagine a body located inside of a massive sphere filled with members
of Metagalaxy. If the body suddenly gets acceleration, then such
acceleration will be similar to the acceleration of the whole Universe
relative to the body. In this process all members of the Universe
participate equally regardless of their distance. If the Universe is
homogeneous and isotropic, then such acceleration for the body means
distortion of local isotropness (rotation) or local homogeneity (uniform
motion). Forces of inertia occur. Mach proposed a viewpoint similar to this
ideology. A system is examined in the Universe [7], in which a big amount of
matter exists on far distances.

Centrifugal forces occur due to real rotation around a real rotation axis,
and thus the local isotropness is distorted in the system of the Universe.
Forces of Coriolis occurring in a rotating system of coordinates are real
forces and are created by the rotation of the whole Universe around the
considered body.

2. Let us discuss a particular case when the metrics is given on a
three-dimensional sphere and is determined by formula:

\begin{equation}
ds^2=r^2\left( \sin ^2\theta d\varphi ^2+d\theta ^2\right)  \label{eq21}
\end{equation}

That is equivalent to the motion of a three-dimensional spherical surface in
the radial direction in the four-dimensional pseudo-Rimman space (18).

\[
\left( 1-\dfrac r{r_k}\right) c^2dt^2-\dfrac{dr^2}{1-r/r_k}=0 
\]

or

\begin{equation}
r=r_k\left( 1-\cos \theta \right) ,\text{ \qquad }r=\dfrac h{mc}\left(
1-\cos \theta \right)  \label{eq22}
\end{equation}

Compton's formula is received for dispersion of light on charged particles
if $\dfrac vc=\cos \theta $, $r_k=\dfrac h{mc}$ are the Compton or inertial
radiuses. In Schwarzschild's solution this particular case brings us to
Einstein's formula [8], which was obtained before the creation of GTR.

\begin{equation}
c=c_0\left( 1-\dfrac{2GM}{c^2r}\right)  \label{eq23}
\end{equation}

The observed Compton effect on charged particles is an expression of the
active inert mass. For big m this effect is not observed due to its small
value.

In the metrics of Schwarzschild when $r\rightarrow r_{gr}$ all processes on
the body in relation to the external observer are ''frozen'', and a collapse
occurs which creates a frozen body, which does not send any signals to the
surrounding environment and interacts with the external world only through
its static gravitational field. Such a formation is called a gravitational
black hall, or a gravitational collapse.

In metrics (18) when $r\rightarrow r_k$, all processes on the body in
relation to the internal observer are frozen. A frozen inert black hall is
formed, which interacts only through its inert filed. The source of the
inert field is the equivalent mass of energy of gravitational field.
Analogously, the source of the gravitational field is the equivalent mass of
energy of the inert field.

Suppose $r\rightarrow r_{gr},r_k$, then the body will be within
gravitational as well as within inertial radiuses. Such situation will occur
only in point $r_0=\left( \dfrac{GM}{c^3}\right) ^{1/2}$, which is
determined by the Plank distance. This point $r_0$ in the metrics of the
four-dimensional pseudo-Rimman space seems to us to be an absolute and
special point. A body determined in such a way may be considered both blind
and deaf.

If gravitational and inert fields are created by spherical bodies, then
their full mass is expressed as

\begin{equation}
m=\dfrac{4\pi }{c^3}\int P_0^0r^2dr  \label{eq24}
\end{equation}

For the gravitational field the tensor is determined by component

\begin{equation}
P_0^0=\dfrac{2r_0}{r^3}  \label{eq25}
\end{equation}

and for the field of inertia

\begin{equation}
\overline{P}_0^0=\dfrac 2{rr_0}  \label{eq26}
\end{equation}

determining the limits of integration. The dead particle with radius $%
r_0=\left( \dfrac{GM}{c^3}\right) ^{1/2}$ and mass does not create any
excitation in the space. The latter remains relativistically flat, although
the space is like being filled up with dead mass (blind and deaf particles).

Such virtual particles are foundations, on which real particles are born by
localizing equivalent energy in the given volume of space through exciting
it. That is why it is assumed that during formation of the particle the
limits of integration are changed from $1$ to $r$ on the one side, and from $%
1$ to $1/r$ on the other. Thus,

\[
m=\int\limits_{1/r}^r\dfrac{2r_0}{r^3}r^2dr\text{ \qquad if \qquad }x=\dfrac{%
r_0}r 
\]

for the source of gravitational field

\begin{equation}
m=\int\limits_{1/x}^x\dfrac{dx}x=2\ln \left| x^2\right|  \label{eq27}
\end{equation}

for the source of the inert field

\begin{equation}
\overline{m}=\int\limits_{1/x}^xxdx=\dfrac 12\left( x^2-\dfrac 1{x^2}\right)
\label{eq28}
\end{equation}

Using the active principle of equivalence again, we receive

\begin{equation}
x^4-1=4\left| x^2\right| \ln \left| x^2\right|  \label{eq29}
\end{equation}

A transcendental algebraic equation is obtained, the roots of which are

\[
\left| x_0\right| ^2=1,\text{ \qquad }x_1=x_0\alpha ^{1/4} 
\]

and their reverse values.

We think that $x_1^4=j$ is the value of Dirack's monopole, if the electric
charge $e^2=1$, then $2j^2=\alpha ^2$ is the value of the constant of the
fine structure.

To conclude, we would like to mention that physical consequences of this
work, including cosmological consequences, are very interesting and will be
discussed in a separate article.

\bigskip

\begin{center}
\textbf{Literature}
\end{center}

1. H.Minkowski, The Principle of Relativity, Dovez Publications, New York
(1952)

2. A. Einstein, Theorie unitaire de champ physique, Ann. Just. H. Poincare,
1. 1-24, (1930)

3. A. Einstein, Ann. Phys., 49, 769 (1916)

4. K. Schwarzshild, Sitrungsber. Preuss. Akad. Wiss., s.424, (1916)

5. D. W.Siama, Roy. Astron. Sos. Monthly Notices, 113, 34 (1953)

6. I. Newton, Philosophiae Naturalis Principia Mathematica, University of
California Press, p. 546 (1966)

7. E. Mach, The Science of Mechanics 2nd ed., Open Court Pable. Co., (1893)

8. A. Einstein, Ann. Phys., 35, 898 (1911)

\end{document}